\documentclass{eas}
\usepackage{graphicx,natbib}
\begin{document}

\TitreGlobal{Mass Profiles and Shapes of Cosmological Structures}

\title{Understanding the Equilibrium Structure of CDM Halos}
\author{Paul R. Shapiro}\address{Department of Astronomy, 
1 University Station, C1400, Austin, TX 78712}
\author{Kyungjin Ahn$^1$}
\author{Marcelo Alvarez$^1$}
\author{Ilian T. Iliev}\address{Canadian Institute for Theoretical
Astrophysics, University of Toronto, 60 St. George Street, Toronto, ON
M5S 3H8, Canada}
\author{Hugo Martel}\address{D\'epartement de physique, de g\'enie 
        physique et d'optique,          
        Universit\'e Laval, Qu\'ebec, QC G1K 7P4, Canada}



%
\runningtitle{Equilibrium Structure of CDM Halos}
\setcounter{page}{23}
\index{Paul R. Shapiro}
\index{Kyungjin Ahn}
\index{Marcelo A. Alvarez}

%
\begin{abstract} 
N-body simulations find a universal structure for the halos which
result from the nonlinear growth of Gaussian-random-noise density
fluctuations in the CDM universe. This talk summarized our attempts to
derive and explain this universal structure by analytical
approximation and simplified models. As an example, we show here that
a 1D spherical infall model involving a fluid approximation derived
from the Boltzmann equation can explain not only the halo density profile
but its phase-space density profile, as well.
\end{abstract}
\maketitle

\section{Introduction}
Numerical N-body simulations of structure formation from
Gaussian-random-noise initial conditions in the CDM universe find a
universal structure for halos. This universality is a fundamental
prediction of the CDM model, but our knowledge is limited to the
``empirical'' N-body results, with little analytical understanding. In
his talk, Shapiro summarized our attempts to fill this gap by a
hierarchy of approximations, each simpler than the last: 1. 3D
gas/N-body simulations of halo formation from simplified initial
conditions; 2. 1D, spherical, analytical models using a fluid dynamics
approximation derived from the Boltzmann equation; 3. An analytical
hydrostatic equilibrium model which follows from the virialization of
top-hat density perturbations.

Most of the work described in that talk is summarized in
Shapiro et al. (2004) and references therein and in \citet{sidm}, with
the exception of our new 
results on the halo phase-space density profile, 
which we shall present here for the first time.
Due to length limitations, we shall limit this paper to just a few
items from category 2 above.  
A more complete version
of Shapiro's talk is available at the meeting website\footnote{http://www2.iap.fr/users/gam/yappa-ng/index.php?album=\%2FIAP05\%2F\&image=Shapiro.pdf}.

\section{Universal Structure of CDM Halos: N-body Results}

CDM N-body halos show universal mass profiles.  The same density
profile fits halos from dwarf galaxies to clusters, independent of
halo mass, of the shape of the density fluctuation power spectrum
$P(k)$, and of background cosmology: $\rho(r)/\rho_{-2}=fcn(r/r_{-2})$,
where $r_{-2}\equiv$ radius where $d\ln\rho/d\ln r=-2$ and
$\rho_{-2}\equiv \rho(r_{-2})$ (e.g. Navarro et al. 2004)\footnote{The
weak mass-dependence suggested by Ricotti (2003) is an exception which
remains to be confirmed.}.  
As
$r\rightarrow \infty$, $\rho\rightarrow r^{-3}$, while as
$r\rightarrow 0$, $\rho\rightarrow r^{-\alpha}$, $1\leq \alpha \leq
1.5$ (e.g. Navarro, Frenk, \& White 1997; Moore et al. 1999).
Diemand, Moore \& Stadel (2004) report that 
\begin{equation}
\rho_{\alpha\beta\gamma}=\frac{\rho_s}
{\left[r/r_s\right]^\gamma
\left[1+(r/r_s)^\alpha\right]^{(\beta-\gamma)/\alpha}}
\end{equation}
with $(\alpha,\beta,\gamma)=(1,3,\gamma)$ summarizes the fits to
current simulations, with $\gamma_{\rm best-fit}=1.16\pm 0.14$.

The profiles of individual halos evolve with time.  The halo mass
grows as $M(a)=M_\infty \exp\left[-Sa_f/a\right]$, where $S\equiv
\left[d\ln M / d\ln a\right](a=a_f)=2$ (Wechsler et al. 2002).  The
density profile concentration parameter, $c=r_{200}/r_s$, also grows,
according to $c(a)=c(a_f)(a/a_f)$ for $a>a_f$ (Bullock et al. 2001;
Wechsler et al. 2002), starting from $c(a) \le 3-4$ for $a\leq a_f$
(initial phase of most rapid mass assembly) (Tasitsiomi et al. 2004).

CDM N-body halos show surprisingly isotropic velocity distributions.
Halos have universal velocity anisotropy profiles, $\beta(r/r_{200})$,
where $\beta(r)\equiv 1-\langle v_t^2\rangle/(2\langle v_r^2\rangle)$,
with $\beta=0$ (isotropic) at $r=0$, gradually rising to $\beta\sim
0.3$ at $r=r_{200}$ (e.g. Carlberg et al. 1997).

CDM N-body halos show universal phase-space density profiles.  N-body
results find $\rho/\sigma_{\rm V}^3\propto r^{-\alpha_{\rm ps}}$, where
$\alpha_{\rm ps}=1.875$ (Taylor \& Navarro 2001), $\alpha_{\rm
ps}=1.95$ (Rasia, Tormen 
and Moscardini 2004), and $\alpha_{\rm ps}=1.9\pm 0.05$ (Ascasibar et
al. 2004). 
(Also related: $P(f)\propto f^{-2.5\pm 0.05}$; Arad, Dekel, \& Klypin 2004).

\section{The Fluid Approximation: 1D Halo Formation From Cosmological
Infall}
The collisionless Boltzmann equation and the Poisson equation can be
used to derive exact dynamical equations for CDM which are identical
to the fluid conservation equations for an ideal gas with adiabatic
index $\gamma=5/3$, if we assume spherical symmetry and a velocity
distribution which is both skewless and isotropic, 
assumptions which approximate the N-body results reasonably well
(Ahn \& Shapiro
2005).  We have used this fluid approximation to show that most of the
universal properties of CDM N-body halos described above can be
understood as the dynamical outcome of continuous cosmological infall.
\subsection{Self-similar gravitational collapse: the spherical infall
model}
In an EdS universe, scale-free, spherically symmetric perturbations
$\delta M/M\propto M^{-\epsilon}$ result in  self-similar structure
formation.  Each spherical mass shell around the center expands until
it reaches a maximum radius $r_{\rm ta}$ and recollapses, $r_{\rm
ta}\propto t^{\xi}$, $\xi=(6\epsilon+2)/(9\epsilon)$.  There are no
characteristic length or time scales besides $r_{\rm ta}$ and Hubble
time $t$.  For cold, unperturbed matter, this results in highly
supersonic infall, terminated by a strong, accretion shock which
thermalizes the kinetic energy of collapse: $r_{\rm shock}(t)\propto r_{\rm
ta}(t)$.  The spherical region bounded by the shock is roughly in
hydrostatic equilibrium, a good model for virialized halos.

\begin{figure}
\centering
\includegraphics[width=6.1cm]{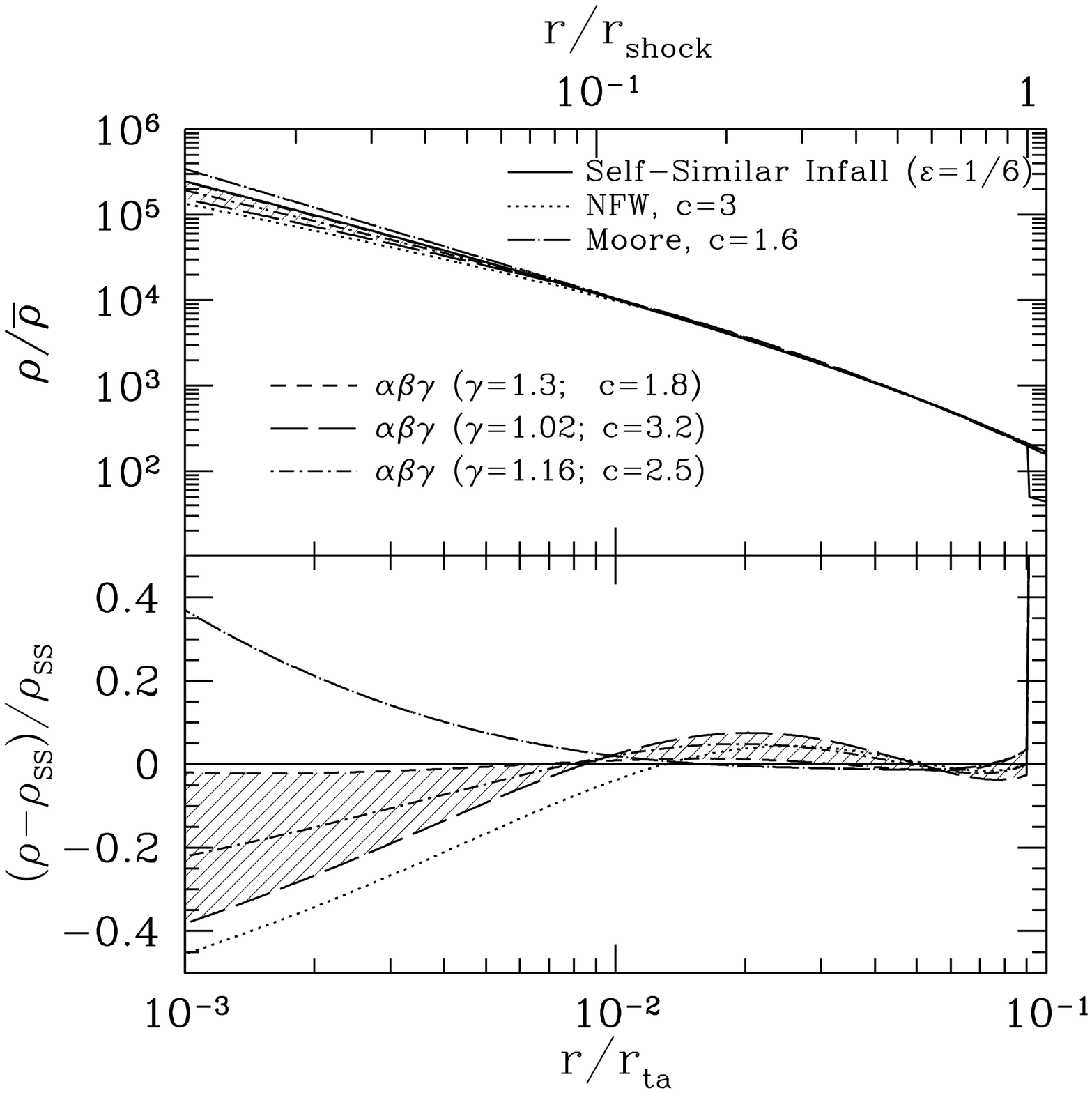}
\includegraphics[width=6.1cm]{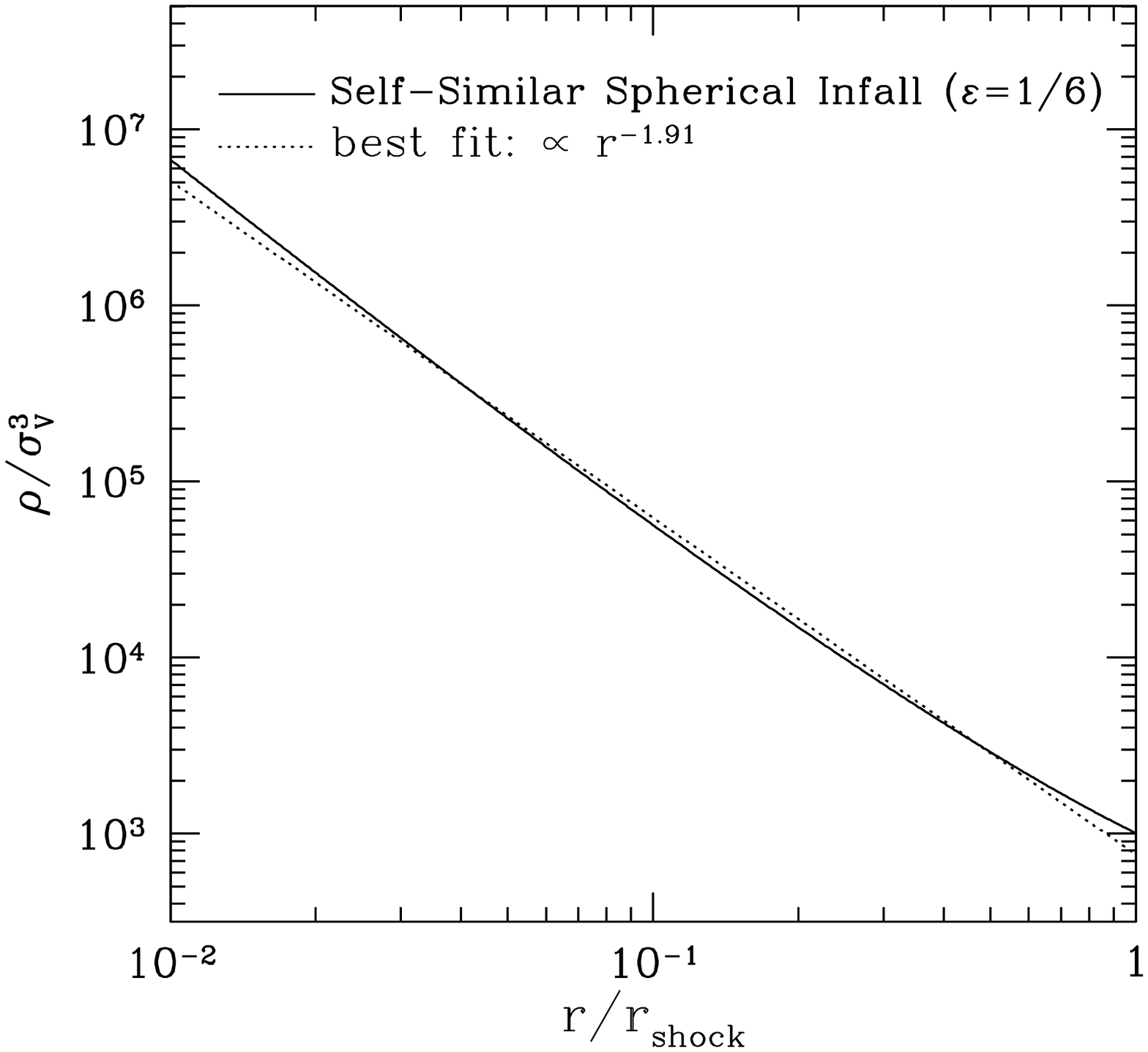}
\caption{
Self-similar Spherical Infall with $\varepsilon=1/6$. (a) (left) (top)
Halo mass density versus radius for analytical,
similarity solution in the
fluid approximation, compared with best-fitting NFW, Moore and $\alpha
\beta \gamma$ profiles with ($\alpha$,$\beta$,$\gamma$)=(1,3,$\gamma$). 
(bottom) fractional deviation of fits from self-similar solution 
$\rho_{\rm SS}$.
(b) (right) Halo phase-space density versus radius for analytical 
similarity solution
compared with best fitting power-law $r^{-1.91}$.}
\end{figure}

Consider halo formation around peaks of the Gaussian random noise
primordial density fluctuations.  If $P(k)\propto k^n$, then
$\nu\sigma$-peaks with $\nu\geq 3$ have simple power-law profiles for
accumulated overdensity inside $r$, $\Delta_0(r)=\delta M/M\propto
r^{-(n+3)}\propto M^{-(n+3)/3}$ (e.g. Hoffman \& Shaham 1985), which
implies self-similar infall with $\epsilon=(n+3)/3$.  For
$\Lambda$CDM, galactic halos are well approximated by $n=-2.5\pm 0.2$
for $10^3\leq M/M_\odot \leq 10^{11}$.  By applying the fluid
approximation to the problem of self-similar spherical infall with
$\epsilon=1/6$ (i.e. $n_{\rm eff}=-2.5$), Ahn \& Shapiro (2005)
derived a 1D, analytical solution for halo formation and evolution, in
which $r_{\rm shock}\propto r_{\rm ta}\propto t^2$, and $M\propto t^4$.  The
resulting self-similar halo density profile inside the accretion shock
agrees with that of CDM N-body halos, with a best-fit
$\alpha\beta\gamma$-profile which has
$(\alpha,\beta,\gamma)=(1,3,1.3)$ (see Figure 1(a)).
As we show in Figure 1(b), this analytical similarity solution for
$\epsilon=1/6$ also derives the universal phase-space density profile
found for CDM N-body halos,
$\rho/\sigma_{\rm V}^3 \propto r^{-\alpha_{\rm ps}}, \alpha_{\rm
ps}\simeq 1.9$. 

\subsection{Non-self-similar infall: Mass Assembly History and the
Evolution of CDM N-body Halo Profiles} 
\begin{figure}
\centering
\includegraphics[width=6.1cm]{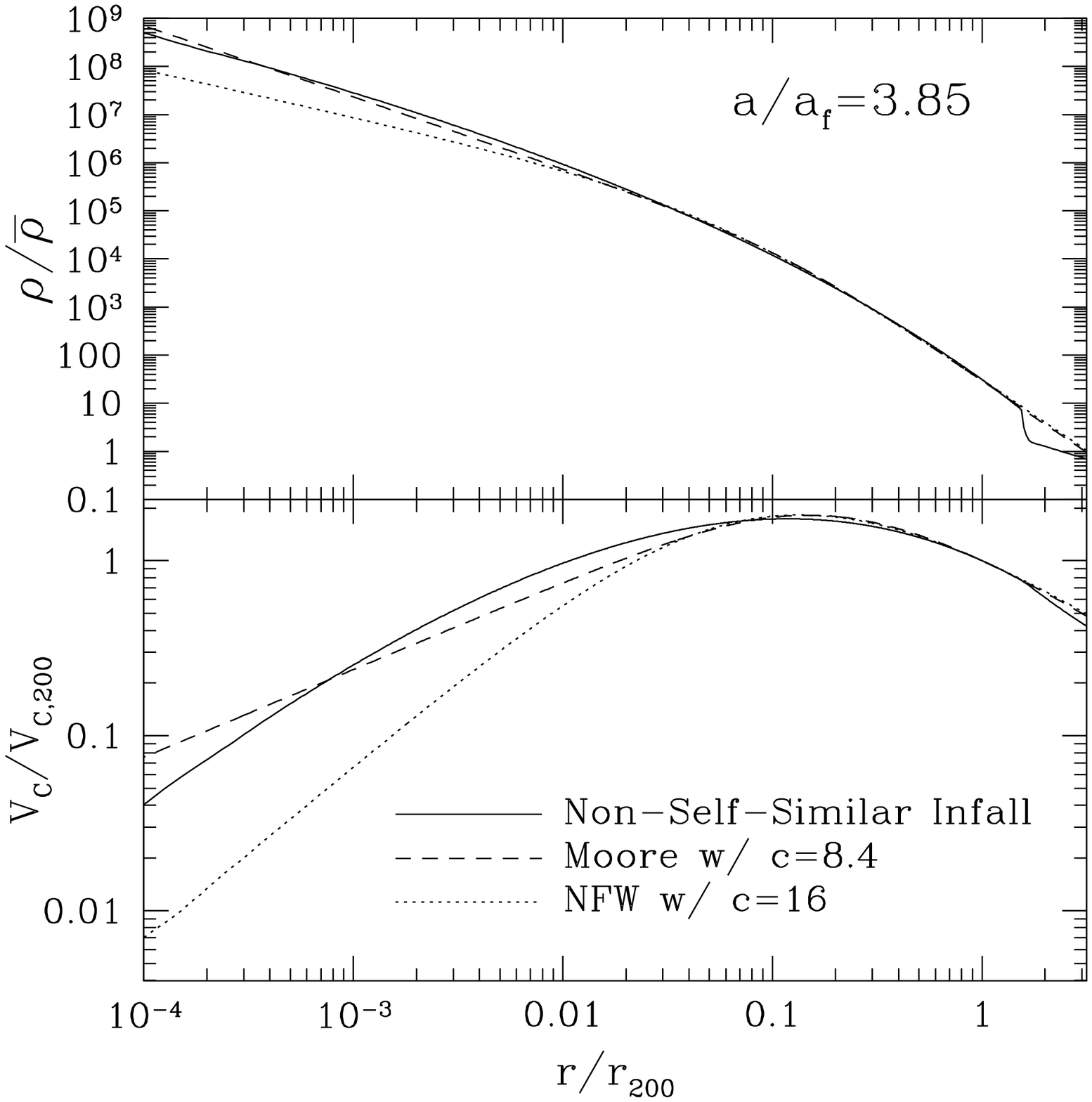}
\includegraphics[width=6.1cm]{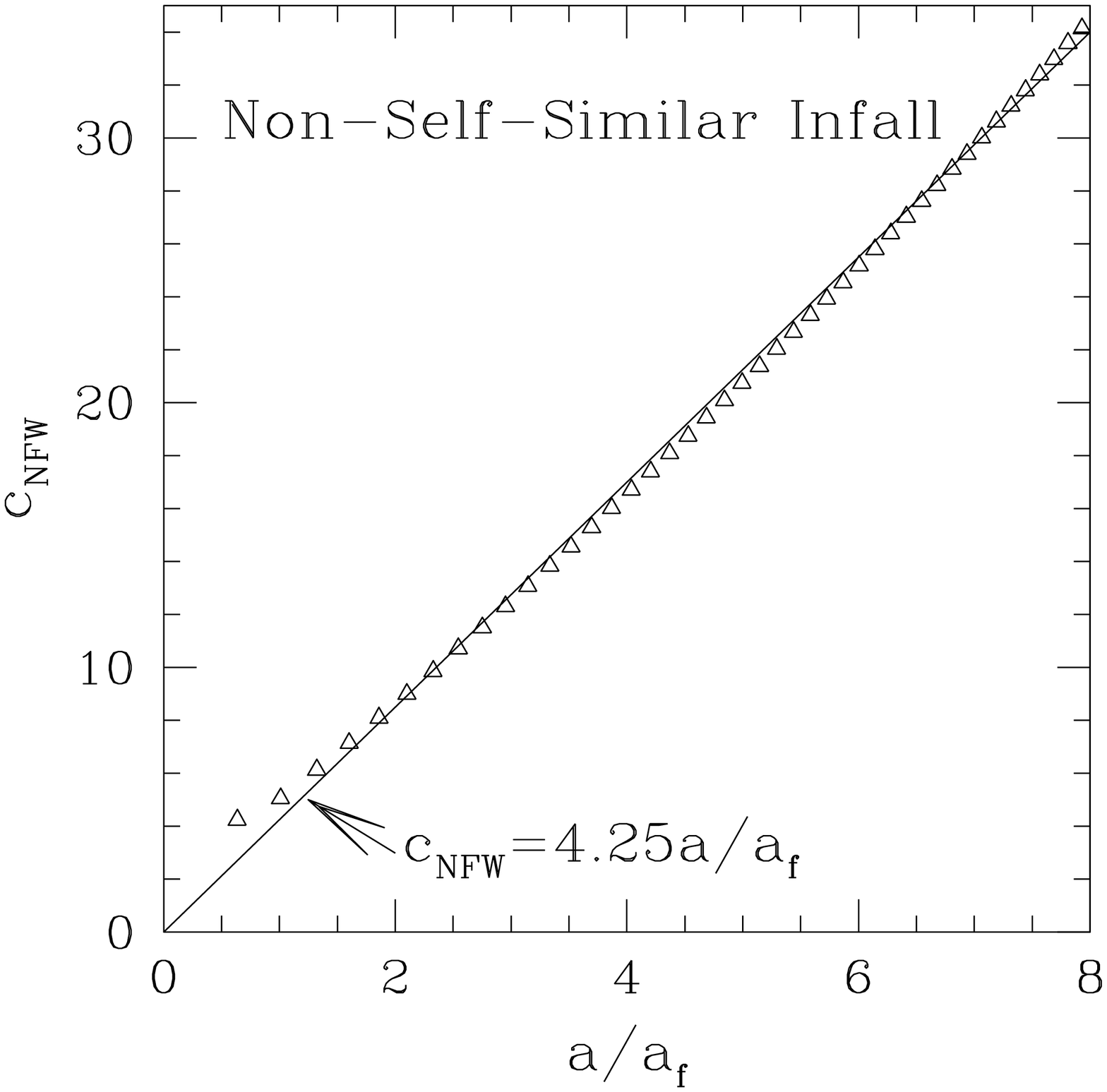}
\caption{
Non-Self-Similar Spherical Infall.
(a) (left) (top) Halo mass density versus radius at epoch $a/a_{\rm
f}=3.85$, according to fluid approximation solution for the
non-self-similar spherical infall rate which makes halo mass grow in
time like Wechsler et al. (2002) fitting formula, compared with
best-fitting NFW and Moore profiles; (bottom) Corresponding halo
rotation curves.
(b) (right) Concentration parameter versus scale factor for the best-fitting
NFW profiles at each time-slice during the evolution of the fluid
approximation solution for time-varying spherical infall at the
Wechsler et al. (2002) rate. }
\end{figure}


\begin{figure}
\centering
\includegraphics[width=8.5cm]{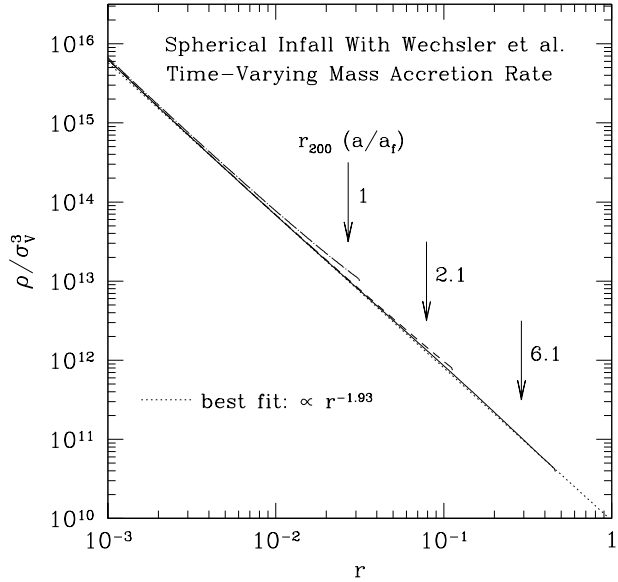}
\caption{Phase-space density profiles for the same non-self-similar
infall solution plotted in Figure 2, for $a/a_{\rm f}=1$, 2.1,
and 6.1, with arrows showing locations of $r_{200}$ at each epoch,
along with best-fitting power-law $r^{-1.93}$.}
\end{figure}
Self-similar infall may provide a good explanation for some halo
properties, but it cannot explain how profile shapes change with
time and depart from self-similarity.  To do this, we have derived the
perturbation profile that leads to the non-self-similar halo mass
growth rate reported for CDM N-body halos by Wechsler et al. (2002)
and used the fluid approximation to derive the halo
properties that result (Alvarez, Ahn, \& Shapiro 2003).  We solved the
fluid approximation equations by a 1D, spherical, Lagrangian hydro code.  These solutions explain
most of the empirical CDM N-body halo properties and their evolution
described above 
as a dynamical consequence of this  time-varying, but
smooth and continuous infall rate.
The halo density profiles which result are well fit by (and
intermediate between) NFW and Moore profiles (over the range
$r/r_{200} \ge 0.01$) (see Figure 2(a)).  These halo density profiles {\em
evolve} just like CDM N-body halos, too.  The halo concentration
parameter grows with time just like CDM N-body halos (see Figure 2(b)).  
In addition, these
solutions yield a halo phase-space density profile, $\rho/\sigma_v^3$,
in remarkable agreement at all times with the universal profile
reported for CDM N-body halos (see Figure 3).  We therefore conclude
that the time-varying mass accretion rate, or equivalently the shape
of the initial density perturbation, is the dominant influence on the
structure of CDM halos, which can be understood simply in the context of
spherical collapse and the accretion of a smoothly-distributed,
spherically-symmetric collisionless fluid.

This work was supported NASA Astrophysical Theory Program grants
NAG5-10825, NAG5-10826, NNG04GI77G, Texas Advanced Research Program
grant 3658-0624-1999, and a Department of Energy Computational Science
Graduate Fellowship to M.A.A.



\end{document}